\documentclass[prl,aps,reprint,graphicx,showpacs]{revtex4-1}
\usepackage{graphicx}
\usepackage{amssymb}
\usepackage{subfigure}
\usepackage{color,soul}
\begin{document}

\title{Two-dimensional electron-gas-like charge transport at magnetic Heusler alloy-SrTiO$_3$ interface}

\author{P. K. Rout,$^{1,\dag}$ Himanshu Pandey,$^{1,\dag}$  Lijun Wu,$^{2}$ Anupam,$^1$ P. C. Joshi,$^1$ Z. Hossain,$^1$ Yimei Zhu,$^{2}$ and R. C. Budhani$^{1,3,\ast}$}
\affiliation{$^1$Condensed Matter - Low Dimensional Systems Laboratory, Department of Physics,
Indian Institute of Technology Kanpur, Kanpur - 208016, India\\
$^2$Brookhaven National Laboratory, Upton, NY - 11973, USA\\
$^3$CSIR-National Physical Laboratory, New Delhi - 110012, India}

\collaboration{$^{\dag}$P.K.R. and H.P. contributed equally to this research.}
\email{rcb@iitk.ac.in, rcb@nplindia.org}
\date{\today}

\begin{abstract}

We report remarkably low residual resistivity, giant residual resistivity ratio, free-electron-like Hall resistivity and high mobility ($\approx$ 10$^4$ cm$^2$V$^{-1}$s$^{-1}$) charge transport in epitaxial films of Co$_2$MnSi and Co$_2$FeSi grown on (001) SrTiO$_3$. This unusual behavior is not observed in films deposited on other cubic oxide substrates of comparable lattice parameters. The scaling of the resistivity with thickness of the films allow extraction of interface conductance, which can be attributed to a layer of oxygen vacancies confined within 1.9 nm of the interface as revealed by atomically resolved electron microscopy and spectroscopy. The high mobility transport observed here at the interface of a fully spin polarized metal is potentially important for spintronics applications.
\end{abstract}

\maketitle

The seemingly extraordinary electronic transport observed in epitaxial films of LaAlO$_3$ (LAO), LaTiO$_3$ and related perovskites grown on TiO$_2$ terminated (001) SrTiO$_3$ (STO) has taken central stage in condensed matter physics research in recent years \cite{Ohtomo,Rastogi,Hotta,Okamoto}. The origin of two-dimensional electron gas (2DEG), whose mobility and carrier density depend strongly on growth temperature and oxygen partial pressure \cite{Ohtomo,Rastogi,Hotta}, and which can be modified further by ultraviolet light \cite{Rastogi} and electric field \cite{Thiel}, has been attributed to interfacial factors such as atomic relaxation, electronic reconstruction, cation intermixing and/or creation of oxygen vacancies \cite{Hwang}. The electronic properties of such interfaces have been studied extensively owing to unusual charge transport \cite{Ohtomo,Herranz}, magnetism \cite{Brinkman,Li,Dikin}, two-dimensional superconductivity \cite{Reyren,Biscaras1,Biscaras2}, and quantum oscillation in the conductivity \cite{Shalom,Caviglia}. While several types of oxides overlayers show unusually large interfacial conductivity \cite{Chen1,Chen2}, the common denominator in all these cases is STO, which even without any overlayer but subjected to subtle surface treatments, can show fascinating 2D electronic behavior \cite{Reagor,Lee}.

Departing from the commonly used approach of growing oxide overlayers, here we show, for the first time, a similar electronic transport realized at the interface of a half metallic Heusler alloy and STO. The Heusler compounds have generated considerable interest in recent years due to a myriad of properties encompassing half metallicity, shape memory effect, thermoelectricity,  superconductivity and topologically inhibited conducting states \cite{Graf}. While our discovery of a highly conducting interface between Heusler alloys and STO can have potential technological applications, the fundamental mechanism for the origin of such a state brings into question the several interpretations given for 2DEG previously.

The thin films of Co-based full-Heusler alloys such as Co$_2$FeSi (CFS) and Co$_2$MnSi (CMS) have been grown on variety of semiconductors and oxide dielectrics\cite{Kasahara,Wang,Schneider,Pandey,Anupam}. The substrates used in present study were (001) LAO, MgO, NGO and STO, whose face diagonal matches quite well with the lattice parameter ($\approx$ 0.565 nm) of CM(F)S. We have deposited a large number (over 70) of highly ordered single phase thin films of CM(F)S under various growth environments using pulsed laser ablation technique \cite{Supple}. The growth rate of 0.0065 nm per laser pulse allows the synthesis of smooth and uniform epitaxial films.

We begin by showing the most striking result, which compares the $\rho$($T$) of CMS films grown on LAO, MgO, NdGaO$_3$ (NGO), and STO in Fig. 1(a). The resistivity of the films on LAO, MgO, and NGO falls by only 25$\%$ of its value at 300 K as we approach 5 K. Contrary to this, the film on STO has a very low residual resistivity ($\rho_0$ $\approx$ 0.08 $\mu\Omega$cm); a parameter which gives a measure of the electron scattering due to defects and impurities present in the system. Furthermore, the films on STO display giant values ($\approx$ 1680) of residual resistivity ratio (RRR). In comparison, the lowest reported $\rho_0$ so far for CMS films on any other substrate is $\approx$ 16 $\mu\Omega$cm \cite{Raphael}. For single crystals, the $\rho_0$ is in the range of 1.5-3.0 $\mu\Omega$cm \cite{Blum,Raphael}. Similarly, the best RRR reported for films and single crystals is limited to only 5-6 \cite{Blum,Raphael}. Since STO is prone to reduction in vacuum at elevated temperatures ($>$ 850$^\circ$C), which may render it conducting \cite{Spinelli}, it is important to rule out this possibility during the film growth. A bare (001) STO substrate treated under the same conditions as used for the film growth shows an insulating behavior (sheet resistance $>$ 1 M$\Omega$). Moreover, the maximum temperature to which the substrates were exposed was $\leq$ 600$^\circ$C, where the reduction of STO is highly unlikely. Furthermore, we recover the insulating nature of the STO after etching off the films with dilute HNO$_3$, which shows that the observed transport property is related to the regions near the film-substrate interface. We have compared the resistivity of CFS/STO film with CFS single crystal and LAO/STO 2DEG systems for further insights [Fig. 1(b)]. Clearly, the $\rho_0$ is an order less than that for single crystal while RRR for CFS/STO is comparatively quite large. Although such high RRR is also observed for reduced STO \cite{Spinelli} and highly oxygen deficient LAO/STO \cite{Ohtomo}, the $\rho_0$ in such systems is two orders of magnitude higher. The X-ray diffraction studies on the films shows a high degree of crystallographic ordering with root mean square interface roughness $\leq$ 1 nm \cite{Supple}. Moreover, their saturation magnetic moments are in accordance with the Slater-Pauling rule even for the thinnest film. All these results suggest that the observed charge transport is not of structural or magnetic origin.

\begin{figure}[t]
\begin{center}
\includegraphics [width=8cm]{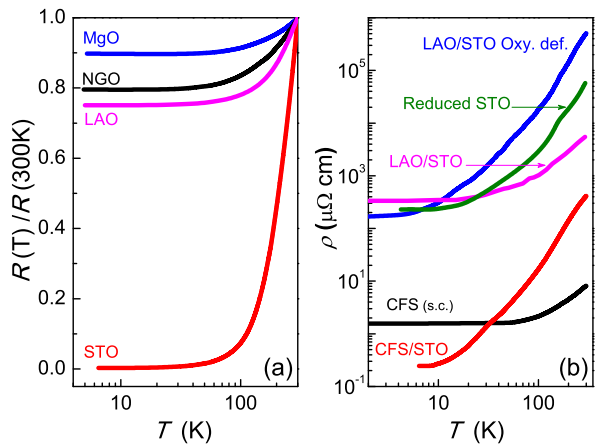}%
\end{center}
\caption{\label{Figure 1} (a) The normalized resistance $R$($T$)/$R$(300 K) of 40 nm thick CFS films on (001) oriented LAO, MgO, NGO, and STO grown in vacuum. (b) The resistivity of LAO/STO \cite{Brinkman}, oxygen deficient LAO/STO \cite{Ohtomo}, reduced STO \cite{Spinelli}, CFS single crystal \cite{Blum} and CFS (12 nm)/STO (Present work). The conducting layer in LAO/STO system is ~10 nm thick while the thickness is ~0.5 mm for the film where the oxygen vacancies dominate the conduction \cite{Supple}.}
\end{figure}

\begin{figure}[t]
\begin{center}
\includegraphics [width=8.5cm]{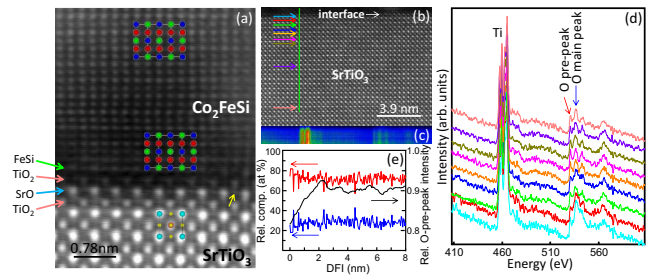}%
\end{center}
\caption{\label{Figure 2} (a,b) STEM image viewed along [110] CFS (or [100] STO) direction, showing the interface between CFS film and STO substrate. The spheres of red, green, blue, cyan, magenta and yellow represent Co, Fe, Si, Sr, Ti and O, respectively. (c) EELS spectra image from the scan line marked by green vertical line in STEM image (b) with a step of 0.05 nm showing Ti \textit{L} edge and O \textit{K} edge. (d) EELS spectra profiles from different positions marked by the arrows in (b). The spectra are normalized with Ti peaks, and each spectrum was averaged and smoothed using the Savitzky-Golay method \cite{Wu}. (e) Relative atomic composition of Ti (blue line) and O (red line) as a function of distance from the interface calculated based on integrated EELS peak intensity in (d) and cross section of Ti and O. The relative atomic percentage of Ti and O is almost 1:3 and remains the same through out STO substrate. The relative O pre-peak intensity using a window of 3.4 eV is also plotted.}
\end{figure}

We believe that the explanation for this extraordinary electronic transport lies in the physics and chemistry of the film-substrate interface, which has been examined by scanning transmission electron  microscopy (STEM) imaging (Fig. 2). The CFS/STO interface as shown in Fig. 2(a) is sharp and coherent with FeSi layer of CFS connecting with TiO$_2$ layer of STO substrate. Interestingly, the contrast of the FeSi layer at the interface (indicated by the green arrow) is slightly stronger than that in the film implying a higher electron density at this layer, which may be due to some Sr at the Fe sites. Moreover, the contrast of the TiO$_2$ and SrO layers at the interface (indicated by the magenta and cyan arrows) is weaker than that in the substrate, suggesting that some Fe and Si may have diffused to these layers. The visible contrast in the O site of SrO layer at the interface (indicated by the yellow arrow) also suggests replacement of O by Si at this site. Based on these observations, we infer a 0.78 nm thick region of inter-diffusion at the interface, which may lead to the chemical doping of STO near the interface. However, the conductivity of STO doped with the 3$d$-transition metal elements is quite low as compared to the conductivity value of $\sim$ 10$^3$ $\Omega^{-1}$cm$^{-1}$ at 300 K observed for CFS/STO films. For example, the 0.1 weight $\%$ Fe doped STO has a conductivity of $\approx$ 2$\times$10$^{-6}$ $\Omega^{-1}$cm$^{-1}$ at 300 K \cite{Blanc}. The electron energy loss spectra (EELS) as shown in Fig. 2(c,d) reveal a significant difference in the intensity of the O-pre-edge (528 eV) relative to that of the O-main-peak (535 eV) as a function of the distance from the interface (DFI), which indicate transitions from O 1$s$ to unoccupied 2$p$ states, and hybridized with Ti-3$d$ states \cite{Zhang}. While the intensity of the O pre-peaks at and near the interface (light blue to blue lines) is low, it increases when the DFI is larger than 1.9 nm (from orange lines). The relative O-pre-peak intensity, defined as the intensity of O-pre-peak divided by that of O-main-peak, gradually increased until DFI = 1.9 nm and then becomes flat, which is suggestive of hole depletion at the interface [Fig. 2(e)]. This interfacial oxygen deficient region of thickness $\approx$ 1.9 nm can lead to the formation of 2DEG at CFS/STO interface. The oxygen vacancies in STO can be created due to interfacial redox reactions with the metallic components of CM(F)S layer. Such effects have been speculated at the interface between STO and other complex oxides with Al, Ti, Zr and Hf elements \cite{Chen1}. However, the oxidation of Fe grown on (001) STO is only observed above 800$^\circ$C \cite{Fu}. We expect a similar threshold for the oxidation of other 3\textit{d} transition metals like Co and Mn. This only leaves the possibility of the redox reactions by Si. To verify such an effect, we have grown thin films of yet another Heusler compound Co$_2$FeAl on STO under the same conditions used for the growth of CF(M)S/STO. These samples show a RRR of only 1-2 and suggest that the Si is responsible for the formation of oxygen deficient region confined at the interface with attendant high mobility electron gas.

\begin{figure}[b]
\begin{center}
\includegraphics [width=8.5 cm]{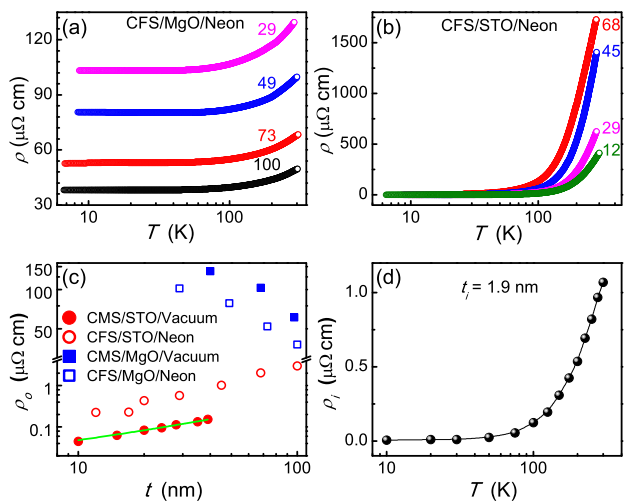}%
\end{center}
\caption{\label{Figure 3} The $\rho$($T$) at different thicknesses (in nm) of CFS deposited on (a) MgO and (b) STO. (c) The $\rho_0$ as a function of thickness. The values of $\rho_0$ are obtained from the fitting of $\rho$($T$) data \cite{Supple}. The line shows the fit according to Eq. (1). ($\bf{D}$) The $\rho$($T$) of the interfacial layer calculated using Eq. (1) for $t_i$ = 1.9 nm.}
\end{figure}

The possibility of such 2DEG is further strengthened from transport measurements on a series of films of varying thickness deposited on MgO and STO (Fig. 3). One would anticipate that the resistivity of thinner film should be greater than that of the thicker film due to enhanced surface scattering, strain induced defects and/or the presence of electrically dead layers at the interface. Indeed, we observe such behavior in the films grown on MgO [Fig. 3(a)]. On the contrary, the $\rho$($T$) of the films on STO reduces with decreasing thickness [Fig. 3(b)], suggesting the presence of an electrically more conducting layer at the interface. A linear extrapolation of the $\rho_0$ vs. $t$ data [Fig. 3(c)] to $t$ = 0 leads to the limiting resistivity $\approx$ 0.008 $\mu\Omega$cm of the interface at 10 K. We have estimated the conductivity of the interfacial layer in the framework of a simple parallel resistor model, which assumes an interfacial layer of thickness $t_i$ and the film with thickness ($t$) while their respective conductivities are $\sigma_i$ and $\sigma_f$. The net effective conductivity is expressed as:
\begin{eqnarray}
\sigma = \sigma_{f} + \frac{t_i\sigma_{i}}{t}
\end{eqnarray}
The $\sigma_{i}$ can be estimated from the slope of the $\sigma$ vs (1/$t$) curve if $t_i$ is known. Assuming the interfacial layer to be 1.9 nm as estimated before, we get a $\rho_{i} \approx$ 0.006 $\mu\Omega$cm at 10 K [Fig. 3(d)], which corresponds to sheet conductance of $G_s \approx$ 32 $\Omega^{-1}$. In comparison, the reported values of $G_s$ for oxide interfaces are order of magnitude lower. For example, a conductance of  $\sim$ 10$^{-3}$ $\Omega^{-1}$ has been reported for LAO/STO \cite{Thiel,Copie} and LaTiO$_3$ (LTO)/STO \cite{Biscaras1}.

\begin{figure}[b]
\begin{center}
\includegraphics [width=8.5cm]{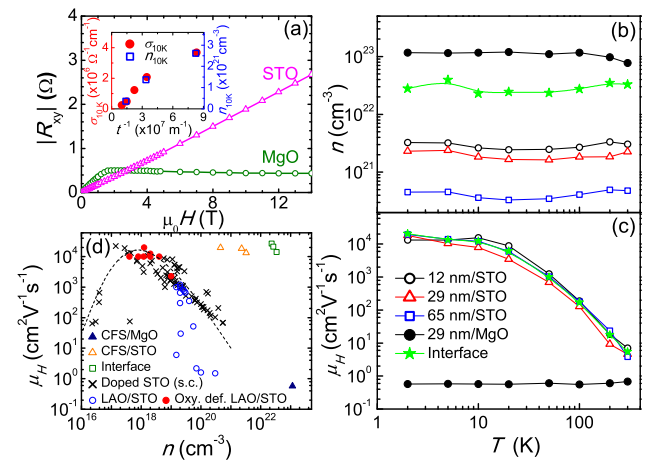}%
\end{center}
\caption{\label{Figure 4} (a) The $|R_{xy}|$($H$) for CFS (12 nm)/STO and CFS(29 nm)/MgO films. The inset shows similarity between the thickness dependent $\sigma$ and $n$ data. The $n$ (b) and $\mu_H$ (c) for 12, 29, and 65 nm thick CFS/STO and CFS (29 nm)/MgO as a function of temperature. All the films are grown in neon environment. The estimated values of $n$ and $\mu_H$ of the interfacial layer are also shown. (d) The $n$ and $\mu_H$ data for single crystal STO doped with oxygen vacancy, Nb, and La are shown by cross symbols \cite{Supple}. The dotted bell shaped curve is guide to the eye. Most of $\mu_H$-$n$ data of LAO/STO can be matched with the data for doped STO assuming a bulk conduction in STO for oxygen deficient LAO/STO films while a two-dimensional conduction through 10 nm thick interface for the rest \cite{Supple}. It also shows our data at 2 K for CFS films on MgO and STO as well as the estimated values corresponding to only interface.}
\end{figure}

The Hall resistance ($R_{xy}$) of CFS film on MgO [Fig. 4(a)] displays a characteristics anomalous behavior of a ferromagnet indicated by a sharp rise of $R_{xy}$ till magnetic saturation. In contrast, for the films on STO, we observe a linear field dependence of $R_{xy}$ up to 14 T with no anomalous contribution. Figure 4(b,c) show the carrier concentration ($n$) and Hall mobility ($\mu_H$) of CFS/STO films. With decreasing thickness, a monotonic increase in $n$ is observed with a highest value of 3.3$\times$10$^{21}$ cm$^{-3}$ at 2 K for 12 nm film, which can be qualitatively explained by considering the parallel resistor model, where
\begin{eqnarray}
n = \frac{{\sigma ^2 }}{{e\left[ {\sigma _f \mu _f  + \frac{{t_i }}{t}\sigma _i \mu _i } \right]}}
\end{eqnarray}
Here, $\mu _i$ and $\mu _f$ are the mobilities of the interfacial layer and the film, respectively. We can assume that $\sigma_f$, $\sigma_i$, $\mu _f$, and $\mu _i$ are independent of $t$ at a particular temperature. Under these conditions and $\sigma \sim$ 1/$t$ [from Eq. (1)], we obtain $n \sim$ 1/$t$, i.e. $n$ increases with decreasing thickness. The thickness independence of$\mu$ [Fig. 4(b)] implies that (1) $n \propto \sigma$ (or 1/$\rho$), which can be seen in the inset of Fig. 4(a), and (2) the scattering due to impurities or defects does not play a dominant role in these films. We have estimated the electronic mean free path ($l$) to be 2-5 $\mu$m at 2 K from the observed values of $\mu_H$ and thus a very large Ioffe-Regel parameter ($k_Fl \sim 12000-22000$). The $n$ is almost independent of temperature for each thickness [Fig. 4(b)]. This is in contrast to the result in LAO/STO system, where $n$ decreases with decreasing temperature due to the carrier freezing at impurity sites \cite{Huijben}. On the other hand, the $\mu_H$ decreases drastically with temperature from a very large value of $\sim$ 20000 cm$^2$V$^{-1}$s$^{-1}$ at 2 K. This indicates that the observed low $\rho_0$ and high RRR are due to a change in $\mu_H$ rather than in $n$. The $n \approx$ 1.2$\times$10$^{23}$ cm$^{-3}$ and $\mu_H \approx$ 0.68 cm$^2$V$^{-1}$s$^{-1}$ of CFS (30 nm)/MgO film are comparable to earlier reports \cite{Schneider2}. Another interesting feature of these data is relatively low values of $n$ for CFS/STO as compared to that for CFS/MgO [See Fig. 4(b)]. The high $n$ in the films on MgO has been attributed to the partially compensated Hall voltage by electron-like and hole-like portions of Fermi surface \cite{Schneider2}. Thus a low $n$ in CFS/STO implies a comparatively high $R_{xy}$ and thus an increase in electron-like portions of Fermi surface, which can be an effect of the interface with higher electron density. However, this simple model needs to be augmented by the consideration of multi-sheeted structure of the Fermi surface as well as effective mass tensors of individual bands, which is still missing in the literature. Furthermore, the parallel resistor model described earlier provides an estimation of $n_i$ and $\mu_i$ [See Fig. 4(b,c)]. From Eq. (2), we have $\mu _i  = \mu  + (\mu  - \mu _f) t \sigma_f/t_i \sigma _i$ and $n_i  = \sigma _i /e\mu _i$ \cite{Supple}. We can see that $n_i$ is greater than $n$ [Fig. 4(b)]. Thus the sum of the carriers of the interface layer and the film will not be equal to total number of carriers extracted from Hall measurements as the mobilities of each component play an important role in such cases. Similar behavior is observed in LTO/STO 2DEG systems, where the electronic transport is governed by two kinds of carriers with different mobilities. Figure 4(c) also shows the $\mu_i$, which almost coincides with those of CFS/STO films. Clearly the interface is solely responsible for the high mobility observed in electronic transport of the films on STO.

To have a better comparison with other conducting oxide interfaces and doped oxides, we show the $\mu_H$-$n$ data taken from literature along with our own CFS/STO values [Fig. 4(d)]. Assuming a 2D or bulk conduction, the data for pure oxide based samples follow a unique bell shaped curve. However, our data for CFS/STO do not follow this behavior. While the $\mu_H$ is of same order as that for oxygen deficient LAO/STO films, the $n$ is at least three orders of magnitude higher. All these results indicates that the formation of quantum well states in STO due to the interface barrier may be at work in addition to oxygen vacancies to have such a high $n$ \cite{Biscaras2,Copie}. In such a scenario, the transfer of electrons from Heusler alloy side to the interfacial STO side can occur leading to higher $n_i$ and these electrons will be highly mobile due to weakening of the charge screening in STO near the interface. We believe that the electronic correlations also play a crucial role in these systems. Using the literature data of $\epsilon \approx$ 330 at 300 K (and 24123 at 4.2 K) for STO \cite{Neville} and the effective mass $m^* \approx$ 3$m_e$ \cite{Dubroka}, where $m_e$ is the electronic mass, we obtain the Bohr radius $a_B \approx$ 5.8 nm at 300 K (and 425 nm at 4.2 K). These values are quite large compared to electron-electron separation ($\sim$ 1/$n_i^{1/3}$) of $\approx$ 0.5 nm, which suggests the electron interactions are important for understanding of these systems.

In summary, we have observed extraordinary electron transport in epitaxial Co$_2$MSi (M = Mn and Fe) films on (001) SrTiO$_3$ with a low $\rho_0$, which is at least an order of magnitude smaller than the values reported in these compounds so far. The films show an giant RRR of $\approx$ 1680 and $\mu_H$ as high as $\sim$ 20000 cm$^2$V$^{-1}$s$^{-1}$. The STEM combined with EELS shows the presence of oxygen deficient region confined within 1.9 nm thickness of STO near the interface, where a two dimensional high mobility electron gas appears to prevail. The thickness dependent study further establishes the presence of an electrically more conducting interfacial layer. The highly spin polarized character of electrons in Heusler alloys adds a magnetic dimension to the problem, which is potentially important for spintronics. Our results are expected to trigger research on the interfaces of several other intermetallics with SrTiO$_3$.

The authors thank Hari Kishan, V. P. S. Awana, M. Shivkumar and the staff of Nanosciences Centre-IIT Kanpur for help in various measurements. Our thanks are also due to S. S. P. Parkin and Michael Pepper for valuable discussions. P.K.R. and H.P. acknowledge financial support from the CSIR, India. R.C.B. acknowledges the J. C. Bose Fellowship of DST.

\end{document}